\documentclass[twocolumn,floatfix,pra,aps,showpacs,superscriptaddress]{revtex4}
\usepackage{epsfig,graphicx}
\usepackage{flafter}
\usepackage{amsmath}
\usepackage{color}
\usepackage{bm}
\usepackage{amssymb}
\usepackage{epstopdf}
\usepackage{amstext}
\usepackage{amsthm}
\usepackage{amsfonts}
\usepackage{latexsym}
\usepackage{multirow}
\usepackage{mathtools}
\usepackage{bm}
\usepackage{bbm}

\begin{document}

\title{
Magnetic defects in an unbalanced mixture of two Bose-Einstein condensates}

\author{A. Gallem\'{\i}}
\affiliation{INO-CNR BEC Center and Dipartimento di Fisica, Universit\`a degli Studi di Trento, 38123 Povo, Italy}
\author{L. P. Pitaevskii}
\affiliation{INO-CNR BEC Center and Dipartimento di Fisica, Universit\`a degli Studi di Trento, 38123 Povo, Italy}
\affiliation{Kapitza Institute for Physical Problems RAS, Kosygina 2, 119334 Moscow, Russia}
\author{S. Stringari}
\affiliation{INO-CNR BEC Center and Dipartimento di Fisica, Universit\`a degli Studi di Trento, 38123 Povo, Italy}
\author{A. Recati}
\affiliation{INO-CNR BEC Center and Dipartimento di Fisica, Universit\`a degli Studi di Trento, 38123 Povo, Italy}

\date{\today}

\begin{abstract}
When the spectrum of magnetic excitations of a quantum mixture is much softer than the density spectrum, the system 
becomes effectively incompressible and can host magnetic defects. These are  characterized by the presence of a topological 
defect in one of the two  species and by a local modification of the density in the second one,  the total density 
being practically  unaffected. For miscible mixtures  interacting with equal intraspecies coupling constants the width 
of these magnetic defects is fixed by the difference $\delta g$ between the intraspecies and interspecies coupling 
constants and becomes larger and larger as one approaches the demixing transition at $\delta g = 0$. When the density 
of the filling component decreases, the incompressibility condition breaks down and we predict the existence of a 
critical filling, below which all the atoms of the minority component remain bound in the core of the topological 
defect. Applications to the sodium atomic spin species \mbox{$\vert F$=$1, m_F$=$\pm1\rangle$} both in uniform and 
harmonically trapped configurations are considered and a protocol to produce experimentally these  defects is discussed. 
The case of binary mixtures interacting with unequal intraspecies forces and experiencing buoyancy is also addressed.
\end{abstract}

\pacs{03.75.Hh, 03.75.Lm, 03.75.Gg, 67.85.-d}

\maketitle

\section{Introduction}

Solitons and vortices are paradigmatic localised excitations inherent of nonlinear systems of different branches, such 
as classical fluids, fiber optics, polyacetylene, or magnets. Solitons \cite{Frantzeskakis2010}, due to the interplay 
between nonlinearity and dispersion, propagate without losing their shape, even after a two-soliton collision. Vortices 
\cite{Fetter2009}, due to the single-valuedness of the order parameter, have quantized circulation, where the quantization 
number is the so-called winding number or the vortex charge.

Among the different physical systems that can be experimentally accessed, ultra-cold atomic gases provide a prominent 
platform for the investigation of solitons and vortices \cite{Dum1998}. On one hand they can be engineered by phase 
imprinting \cite{Denschlag2000}, density imprinting, quantum quenches and on the other hand, they can provide important 
information on the superfluidity of the gas. Soon after the realization of Bose-Einstein condensation, different kind of 
solitons and vortices have been experimentally observed \cite{Burger2002,Matthews1999,Madison2000}. 

In this paper, we study the nature of vortices and solitons in a Bose-Einstein condensate interacting with a second condensate. 
Two-component condensates were experimentally achieved a few years after the first experimental observation of Bose-Einstein 
condensation \cite{Myatt1997}. The physics of solitons in these systems has been already the object of extensive theoretical 
work \cite{Ohberg2001,Yan2012b,Busch2001,Yan2011,Nistazakis2008,Sartori2013}. In our work we focus on the regime of equal 
intracomponent coupling constants and assume miscibility. Moreover, we will consider small values of the difference $\delta g$ 
between the intraspecies and interspecies coupling constants so that the magnetic spectrum is much softer than the density 
one. In the following we will refer to this condition as to the {\it incompressibility condition}. In this case, a defect 
in one component will be compensated by a density modulation in the second component in such a way that the total density 
profile is not affected \cite{Qu2016}, the spin density exhibiting instead a highly nonlinear local modification. We refer 
to these objects as to {\it magnetic defects}. One should notice however that in reality, the incompressibility condition 
is not perfectly fulfilled, and as a consequence, the total density is weakly modified in the core region. 

Assuming incompressibility we can derive a single equation for the defects, which resembles the single-component counterpart, 
but with a non-trivial modification of the kinetic part. We also show that in the limit of large unbalance, with the defect 
created in the minority component, one recovers the well-known single-component equations, with a strongly renormalized 
healing length related to the susceptibility of the mixture. In the opposite limit, i.e., when the defect is in the majority 
component, the incompressibility condition cannot be valid. In particular a single impurity atom will be bound to the defect 
of the majority component, which is essentially not affected by its presence. The crossover between the magnetic defect and 
the regime of bound states will be also explicitly addressed. 

As already pointed out there have been already a number of studies concerning solitons and vortices in two-component 
condensates. Topologically speaking, the solitons we are dealing with belong to the same family of the dark-brigth solitons 
first introduced in the cold atom field by Busch and Anglin in Ref. \cite{Busch2001}, and found experimentally a few 
years later in Refs. \cite{Becker2008,Hamner2011}. In the case of vortices, Ref. \cite{Law2010} claims the discovery of 
the vortex-bright soliton, the topological extension of the dark-brigth soliton to the vortical case. More close to our 
perspective, Ref. \cite{Eto2011} (extended by Ref. \cite{Mason2013}), and Ref. \cite{Danaila2016}, attacked the problem 
of two components with a vortex hosted in one of the components. Similar analysis have been also carried out in the spin-1 
Bose-Einstein condensate case (both theoretically \cite{Ji2008,Lovegrove2012,Gautam2017} and experimentally \cite{Seo2015}), 
and also in rotating condensates \cite{Saarikoski2010}. 

In the present work we are interested in systems close to fulfill the incompressiblity condition. Assuming this condition, 
the authors of Ref. \cite{Qu2016} were able to obtain analytical solutions for the case of moving magnetic solitons in 
balanced mixtures. In the present work the incompressibility assumption is tested against the numerical solution of the 
two-component Gross-Pitaevskii equations, both in the homogeneous case and in the presence of a harmonic trap. We show that 
in relevant available experimental regimes, incompressibility is an excellent approximation. We also detail the implication 
of having non-equal intra-species interaction strengths and the appearance of the bouyancy effect in the presence of a 
harmonic trap.

The paper is organised as follows: In Section \ref{sect:homomatter} we will adopt the incompressibility assumption, which 
corresponds to assuming that the total density of the system, differently from the spin density, is not affected by the 
topological defect. This assumption will allow us to derive a variational energy functional that we then used in Section 
\ref{sect:mv_homomatter} to describe the magnetic vortex. In the same section we will also calculate the energy cost associated 
with the magnetic vortex and compare it with the one of a vortex line in a single-component condensate. We will also discuss 
the case when the incompressibility condition breaks down and bound states of the minority component within the quantum defect 
emerge. In Section \ref{sect:harmonic} we carry out numerical solutions of the coupled Gross-Pitaveskii equations in the 
presence of harmonic trapping. Since the incompressibility condition cannot be satisfied when there are too few atoms in the 
second component we study in detail the crossover between bound atoms in the vortex core to the magnetic vortex in Section 
\ref{sect:boundto}. In Section \ref{sect:ms_homomatter}, we focus on the case of a magnetic dark soliton, for which analytical 
expressions for any polarisation can be obtained, extending the work of Ref. \cite{Qu2016}. We present our conclusions in 
Section \ref{sect:conclusions}. We also provide some future perspectives and comment about possible experimental feasibility. 
Finally, in Appendix \ref{sect:bouyancy} we compare the magnetic vortex scenario in a trap with the case of unequal intraspecies 
interactions, which is known to give rise to bouyancy and phase separation between the two components even if the mixture is 
miscible in uniform matter \cite{Matthews1999}.

\section{Magnetic topological defects in homogeneous matter}
\label{sect:homomatter}

We consider a mixture of atomic Bose gases in two different hyperfine levels. The mixture is characterised by two order 
parameters $\Psi_1$ and $\Psi_2$. At the mean-field level, the stationary solutions are obtained by minimising with respect 
to the order parameters the Gross-Pitaevskii (GP) energy functional $E_{\rm GP}=\int dV\;\varepsilon_{\rm GP}$, with the 
energy density given by:
\begin{align} 
\label{GPEnergy}
\varepsilon_{\rm GP}=&\sum_{i=1,2}\left[\frac{\hbar^2}{2m}|\nabla\Psi_i|^2+(V_{\rm ext}-\mu_i)|\Psi_i|^2+g_{ii}|\Psi_i|^4\right]\nonumber\\
&+g_{12}|\Psi_1|^2|\Psi_2|^2\,,
\end{align}
where $m$ is the atomic mass, $\mu_i$ are the chemical potentials and $V_{\rm ext}$ a possible external trapping potential. 
The interaction strengths $g_{ii}=4\pi \hbar^2 a_{ii}/m$ and $g_{12}=4\pi \hbar^2 a_{12}/m$ are given in terms of the intraspecies 
$a_{ii}$ and interspecies $a_{12}$ $s$-wave scattering lengths, respectively. The mixture is stable against phase separation 
as long as \begin{equation}
\delta g\equiv \sqrt{g_{11}g_{22}}-g_{12}>0.            
           \end{equation}
 As mentioned in the introduction, we are interested in the magnetic aspects 
of solitons and vortices and we assume $a_{11}=a_{22}=a$, i.e., $g_{11}=g_{22}=g$. The condition $g\gg\delta g=g-g_{12}>0$ 
ensures that the total density will be almost unaffected by the presence of the magnetic defect \cite{Qu2016} (incompressibility 
condition). Such a regime can be experimentally realised by using $^{23}$Na in the two hyperfine states \mbox{$\vert F$=$1, m_F$=$\pm1\rangle$} 
for which the scattering lengths are \mbox{$a_{11}=a_{22}=54.54\,a_B$} and \mbox{$a_{12}=50.78\,a_B$}, where $a_B$ is the Bohr 
radius.

Let us first  consider the homogenoues case ($V_{\rm ext}=0$). The presence of a trapping potential will be analyzed in Section 
\ref{sect:harmonic}, but we anticipate here that our conclusions remain valid also in that case, provided the width of the 
defect is much smaller than the size of the atomic cloud. 

Our Ansatz for the topological excitations exploits the incompressibility of the density with respect to the spin channel, i.e. 
we constrain the densities $n_i=\vert\Psi_i\vert^2$ in the variational calculation by asking
\begin{equation}
n=n_1({\bf r})+n_2({\bf r})\,,
\label{eq:incompressibility}
\end{equation}
which is equivalent to set the total density of the magnetic vortex equal to the total density of the ground state. Writing the 
condensate wave-functions as $\Psi_i=\sqrt{n_{i,0}} f_i({\bf r}) e^{i\phi_i({\bf r})}$, $i=1,\;2$, with $n_{i,0}=n_i(r\rightarrow \infty)$ 
the asymptotic values of the condensate densities, the energy functional can be written as
\begin{align}
\label{GPEnergy-const}
&\frac{\varepsilon_{\rm GP}}{4\delta g n_{1,0}^2}=\frac{1}{4}f_1^2(f_1^2-2)+\nonumber\\
&\frac{1}{2}\left[\frac{n (\nabla_{\bm \eta} f_1)^2}{n-n_{1,0}f_1^2
}+f_1^2((\nabla_{\bm \eta}\phi_1)^2-(\nabla_{\bm \eta}\phi_2)^2)+\frac{n}{n_{1,0}}(\nabla_{\bm \eta}\phi_2)^2\right],
\end{align}
where we have absorbed the constant terms in the definition of the energy density and rescaled ${\bf r}\rightarrow {\bm \eta}={\bf r}/\xi_s$ 
by introducing the {\it in-medium} spin healing length 
\begin{equation}
\label{ximedium}
\xi_s=\frac{\hbar}{\sqrt{4m\,\delta g\,n_{1,0}}}\,.
\end{equation}
The meaning of $\xi_s$ is particularly clear when a topological defect is considered only in the component $1$ and we consider 
a vanishing phase for the component 2 ($\phi_2=0$). In the limit $n_{1,0}\ll n_{2,0}$ the healing length $\xi_s$ then provides 
the only length scale of the problem and the energy density (\ref{GPEnergy-const}) reduces to  the energy density of a single 
component condensate with a renormalized value for the healing length. The length scale characterizing the new solution is deeply 
modified with respect to the {\it density healing length} $\xi_d=\hbar/\sqrt{2mg n_0}$ that appears in a single-component vortex 
with the same asymptotic value of the density. In fact $\delta g$ is assumed to be significantly smaller than the coupling constant 
$g$. In general, near the demixing transition, where $\delta g \to 0$, the width of the magnetic vortex core can become significantly 
large. For instance, for the states \mbox{$\vert F$=$1, m_F$=$\pm1\rangle$} of $^{23}$Na one has $\xi_s/\xi_d=\sqrt{g/2\delta g}=2.69$. 
Let us finally notice that $\xi_s$ fixes the spin speed of sound $c_s$ \cite{Pitaevskii2016} of a homogeneous mixture in the limit 
$n_{1,0}\ll n_{2,0}$ via the relation $m c_s \xi_s=\hbar/\sqrt{2}$.

In the opposite limit, $n_{1,0}\gg n_{2,0}$ the incompressibility condition cannot be satisfied. In this limit the defect in 
component $1$ is essentially unaffected by the presence of the minority component that will be trapped forming a bound state.

In the following Section we show numerically that indeed the incompressibility condition is well satisfied for vortices and solitons 
provided $n_{2,0}$ is not too small. For the vortex state we explore in detail the crossover between the magnetic vortex and the 
bound state regime in Section \ref{sect:boundto}.

\section{Magnetic vortices in homogeneous matter}
\label{sect:mv_homomatter}

\begin{figure}[t!]
\includegraphics[width=\linewidth]{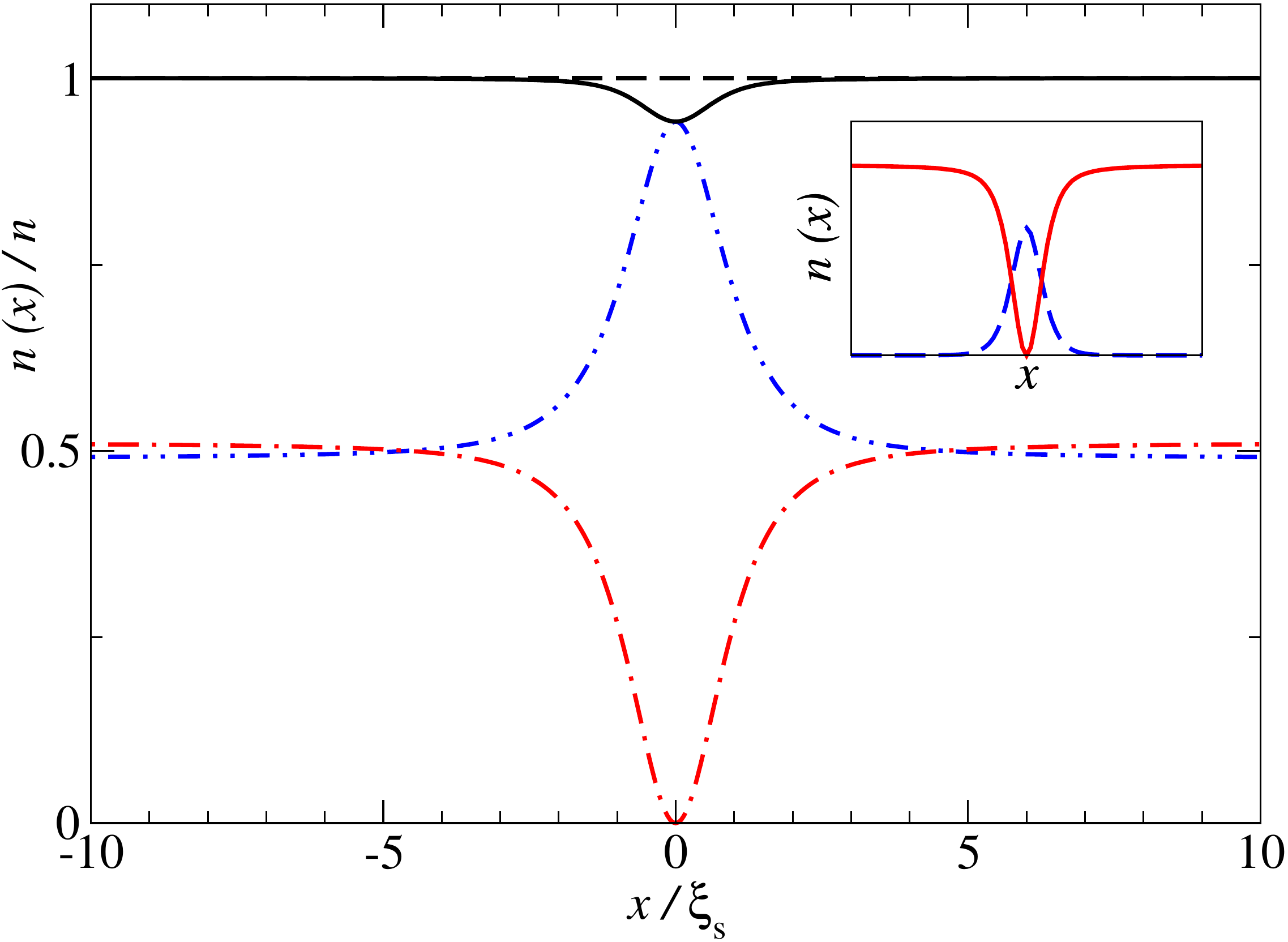}
\caption{The dot-dashed red (dot-dot-dashed blue) line respresents the density (normalized to the total background 
density) along the $x$-axis of component $1$ ($2$) in a magnetic vortex. The solid (dashed) black line shows the 
total density  along the $x$-axis of the magnetic vortex (ground state). \mbox{$a_{11}=a_{22}=54.54\,a_B$} and 
\mbox{$a_{12}=50.78\,a_B$}. The total number of particles is $N=10^5$ and the mass is that of $^{23}$Na. The insets 
pictorically displays the case where the density in the second component is small enough to avoid fulfilling the 
incompressibility condition. In this case, one obtains a bound state in the core of the vortex. } 
\label{fig:homovort}
\end{figure}

In this section we will specialize Eq. (\ref{GPEnergy-const}) to the case in which only component $1$ has a vortex, i.e.,
\begin{equation}
\Psi_1(r,\theta)=\sqrt{n_{1,0}}\,f_1(r)\exp(i\theta)\,,
\end{equation}
and $\phi_2=0$. For the sake of simplicity we study a two-dimensional case with polar coordinates $(r,\theta)$. The 
equation that $f_1$ must satisfy is obtained by imposing $\delta\varepsilon_{\rm GP}/\delta f_1=0$, which leads:
\begin{align}
&\partial_\eta^2 f_1+\left(1-\frac{1}{\eta^2}\right)f_1-f_1^3+\nonumber\\
&\frac{n_{1,0}f_1}{n-n_{1,0}f_1^2}\left[f_1\partial_\eta^2 f_1+\frac{n}{n-n_{1,0}f_1^2}\left(\partial_\eta f_1\right)^2\right]=0.
\label{eq:mvortex_diff_eq}
\end{align}
The first line of Eq. (\ref{eq:mvortex_diff_eq}) is formally the same as the vortex equation for a single component (see 
e.g. \cite{Pitaevskii2016}) with healing length $\xi_s$. The second line is a term that appears due to the presence of 
the second component. This correction vanishes as $n \gg n_{1,0}$, when the incompressibility condition becomes more and 
more accurate. In this limit the vortical solution is then formally identical to the one of a single component condensate 
but with a width, which is fixed by $\xi_s$, increased (see Eq. (\ref{ximedium})) as a consequence of the interaction with 
the second component. We have verified the validity of the incompressibility assumption for $^{23}$Na by numerically solving 
the two coupled Gross-Pitaevskii equations:
\begin{align}
-\frac{\hbar^2}{2m}\nabla^2\Psi_1+g_{11}\vert\Psi_1\vert^2\Psi_1+g_{12}\vert\Psi_2\vert^2\Psi_1&=\mu_1\Psi_1\nonumber\\
-\frac{\hbar^2}{2m}\nabla^2\Psi_2+g_{22}\vert\Psi_2\vert^2\Psi_2+g_{12}\vert\Psi_1\vert^2\Psi_2&=\mu_2\Psi_2\,.
\label{eq:GPE}
\end{align}
The vortical solution is obtained using the imaginary time step method starting from an Ansatz that captures both the phase pattern 
of the wave functions and the increase of the healing length in the case of magnetic defects. The results for the balanced case 
($n_{1,0}=n_{2,0}$) are reported in Fig. \ref{fig:homovort} \footnote{In this figure, we have studied the $2D$ case, by assuming 
a harmonic transverse confinement with $\omega_z=2\pi\times 1.592$ kHz, that has accordingly rescaled the coupling constants. From 
now on, we will assume this transverse confinement. Details about this renormalization can be found in Sect. \ref{sect:harmonic}.}. 
In this case the magnetization \mbox{$n_1-n_2$}, is localized in a small region of the order of $\xi_s$. Indeed we numerically find 
that the healing length of the magnetic vortex is very close to the value (\ref{ximedium}) predicted in the incompressible regime 
(see Fig. \ref{fig:fig7}), a feature which we can prove more explicitly for dark solitons in Section \ref{sect:ms_homomatter}, where 
analytic results are available for all values of $n_1$ and $n_2$. In the case of the magnetic vortex for a balanced mixture the 
numerical solution reveals the occurrence of a small dip  in the total density at the position of the core of the magnetic vortex, 
caused by the finte compressibility of the mixture. We will later show that this dip  disappears as $\delta g\rightarrow 0$, or 
if $n_{2,0}\gg n_{1,0}$.

\subsection*{Energy of magnetic vortices}

The energy of the magnetic vortex can be computed through the energy functional by substracting the ground state 
energy $\varepsilon_{\rm GS}=-\frac{1}{2}gn^2+\delta gn_1n_2$ from the energy of the magnetic vortex. It leads to 
$E_{MV}=\int \varepsilon_{\rm MV} \,d{\bf r}$, where:
\begin{equation}
\varepsilon_{\rm MV}=\frac{\hbar^2n_{1,0} n}{2m}\frac{(\nabla f_1)^2}{n-n_{1,0}f_1^2}+ \frac{\hbar^2n_{1,0}}{2mr^2}f_1^2+\delta g n_{1,0}^2 (f_1^2 -1)^2
\end{equation}
and the integral extends over a disk of radius $R$. The calculation of the energy of the vortex is important because it 
gives access to the value of the rotational frequency required to make the vortical configuration energetically favourable 
\cite{Pitaevskii2016} (see  discussion below). 

In order to evaluate the energy we make use of the approximate Ansatz:
\begin{equation}
f_1(r)=
\begin{cases}
\frac{1}{2}\frac{r}{\xi_s}    & \mbox{if  }  r\leq \sqrt{2}\,\xi_s\\
\sqrt{1-\frac{\xi_s^2}{r^ 2}} & \mbox{if  }  r >   \sqrt{2}\,\xi_s
\end{cases}
\,.
\label{eq:functionc1}
\end{equation}
for the vortex profile which captures the main physics of a single-component vortex line, \footnote{The energy of 
a single vortex computed with this expression yields $L\pi \hbar^2n/m\ln(1.517R/\xi_d)$, which is very accurate, 
taking into account that the correct one, found numerically in Ref. \cite{Ginzburg1958}, gives the same expression 
but with the prefactor $1.46$ instead of $1.517$ in the logarithm.} with a rescaled healing length $\xi_s$. By using 
the Ansatz (\ref{eq:functionc1}) the magnetic vortex energy can be written as 
\begin{align}
E_{MV}=&E_{1V}+\int \frac{\hbar^2n_{1,0}}{2m}\frac{n(\nabla f_1)^2}{n-n_{1,0}f_1^2}\,d{\bf r}=E_{1V}+\nonumber\\
&\frac{\pi\hbar^2n_{1,0}}{m}\left(\left(1-\frac{n}{n_{1,0}}\right)\ln\left(\frac{2n-n_{1,0}}{2n-2n_{1,0}}\right)-\right.\nonumber\\
&\left.\frac{n}{n_{1,0}}\ln\left(1-\frac{n_{1,0}}{2n}\right)\right)\,,
\label{emvhom}
\end{align}
where $E_{1V}=\pi\hbar^2n_{1,0}/m\,\ln\left(1.46\,R/\xi_s\right)$ is the energy of a single component vortex \cite{Pitaevskii2016} 
with a rescaled healing length $\xi_s$. From Eq. (\ref{emvhom}), one can cheack that for $\delta g$ small enough the magnetic 
vortex has a smaller energy cost than the single component one. In the limit $n_{2,0}\rightarrow n$, the correction vanishes 
and the magnetic vortex has a lower energy for any value $\delta g<g$. We will later show that this conclusion is verified 
numerically also in presence of a trap and without imposing explicitly the incompressibility condition.

One can also notice an increasing of the energy when $n_{1,0}\to n$, i.e., $n_{2,0}\to0$, a regime in which the incompressibility 
condition is no longer fulfilled. In this case the system can not be described as a magnetic vortex, but as a single (or a 
few-particle) state bound in the core of the quantum vortex.

\subsection*{Single impurity trapped  in the core of a vortex}
\label{sect:impurity}

In the limit of extreme diluteness of the second component of the mixture, the system must be described as a single particle in an 
effective potential given by the interaction with the density of the majority component which hosts the defect. Therefore from Eq. 
(\ref{eq:GPE}) one can write a Schr\"odinger equation for the wave function $\Psi_2$ of the impurity in the form: 
\begin{equation}
-\frac{\hbar^2}{2m}\nabla^2\Psi_2+ g_{12}\vert\Psi_1\vert^2\Psi_2=\mu_2\Psi_2\,.
\label{EqSPsi2}
\end{equation}
For the order parameter \mbox{$\Psi_1(r,\theta)=\sqrt{n_{1,0}}\,f_1(r)\exp(i\theta)$} we assume the known solution for a single 
component vortex with healing length equal to $\xi_d$, since in this limit $\Psi_1$ is not affected by the interaction with the 
impurity. In this configuration, the impurity sees the vortex core as a trapping potential as shown in the inset of Fig. 
\ref{fig:homovort}.

When we add more atoms of the minority component the width of $\Psi_2$ is enlarged due to the repulsive intraspecies interaction 
and the width of the vortex is enlarged. An important question is what will be the fate of the filling of the vortex core when 
the numbers of atoms of the minority component becomes larger and larger. We will show numerically in Section \ref{sect:boundto} 
that the localised state evolves into the magnetic vortex discussed before. We also derive a simple model to estimate the threshold 
between the two regimes in terms of the atoms of the minority component. Indeed, as long as the number of atoms in component $2$ 
is small, they can be hosted in the vortical region, while after a certain critical number they will diffuse outside the vortex core, 
constituting, at large distances, a uniform gas with density $n_{2,0}$.

\section{Magnetic vortices in a 2D harmonic trap}
\label{sect:harmonic}

The calculation of magnetic vortices in the presence of harmonic trapping is motivated by several reasons. Experimentally, harmonically 
trapped gases are in fact well suited to produce vortical configurations and, consequently, their study represents a topic of primary 
interest. Moreover, the presence of harmonic trapping is particularly useful to investigate the buoyancy effect in the case of unequal 
intra-species interactions, as we will discuss in Appendix \ref{sect:bouyancy}. In the following, we will consider Bose gases hosted 
by an axially symmetric harmonic trap 
$$V_{\rm ext}(r,z)={1\over 2}m(\omega^2 r^2+\omega_z^2 z^2)\,,$$
with $r^2=x^2+y^2$. We also assume \mbox{$\omega_z\gg\omega$}, in such a way that the $z$ degree of freedom is frozen in the ground 
state and a two-dimensional (2D) simulation is enough. We consider in the numerics the parameters for the $^{23}$Na discussed at the 
beginning and renormalise the three-dimensional interaction strength to the 2D values by integrating along $z$, i.e., using in the 
simulation the scattering lengths $a\rightarrow a/\sqrt{2 \pi} \,l_z$ with $l_z=\sqrt{\hbar/m\omega_z}$.

\begin{figure}[t!]
\includegraphics[width=\linewidth]{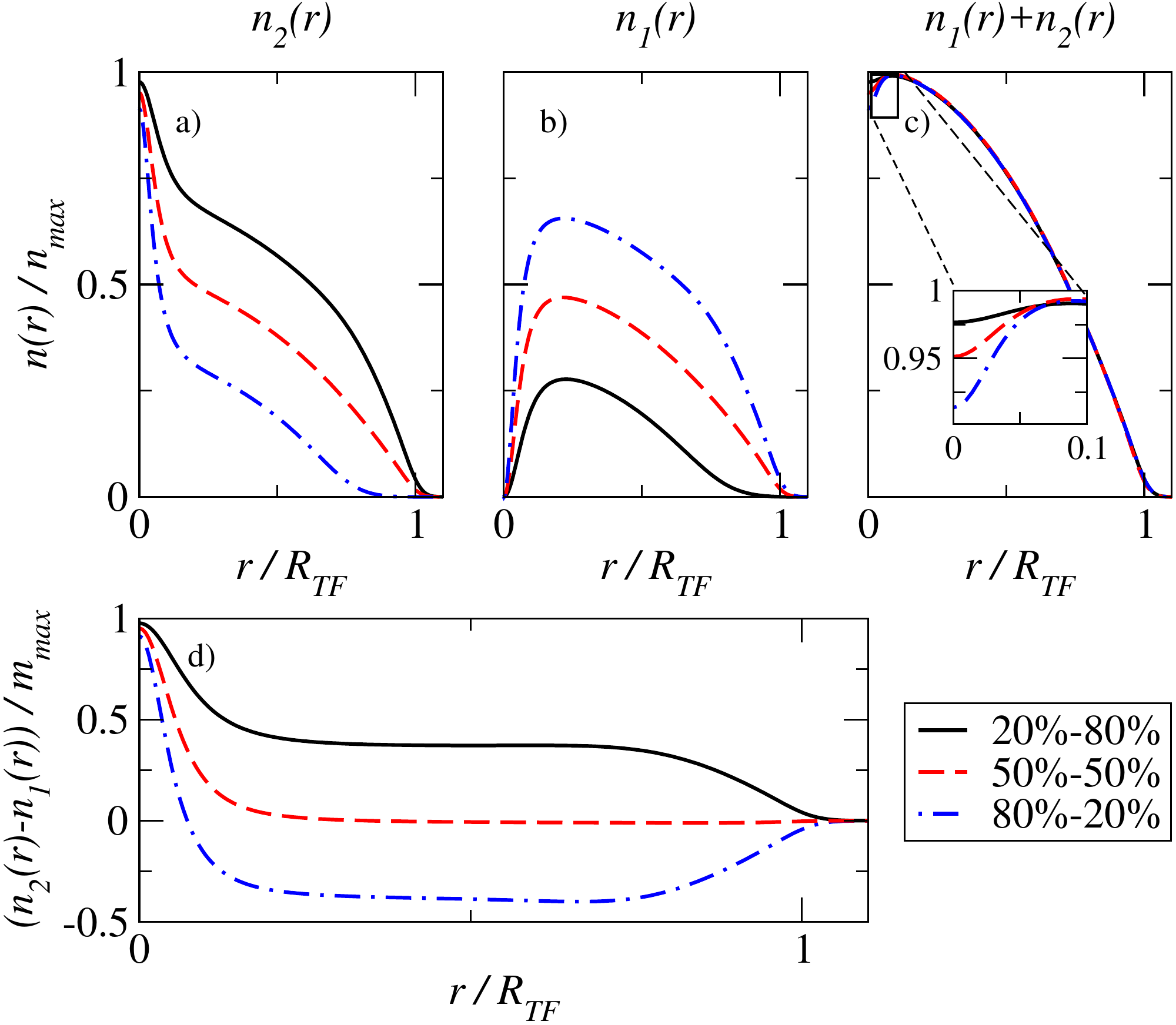}
\caption{a) Density along the radial axis of component $2$ for different values of the population of each 
component: \mbox{$N_1=2\times10^4$} and \mbox{$N_2=8\times10^4$} (solid black line), \mbox{$N_1=5\times10^4$} 
and \mbox{$N_2=5\times10^4$} (dashed red line), \mbox{$N_1=8\times10^4$} and \mbox{$N_2=2\times10^4$} 
(dot-dashed blue line). Panels b), c) and d) represent density of component $1$, the total density and 
the spin magnetization \mbox{$n_2-n_1$}, respectively, following the same legends for the curves. The 
rest of the parameters of the system are: \mbox{$a_{11}=a_{22}=54.54\,a_B$}, $N=10^5$ (total number of 
$^{23}$Na atoms), and $\omega=2\pi\times15.92$ Hz. The densities are written in harmonic oscillator units. }
\label{fig:fig2}
\end{figure}

The results are shown in Fig \ref{fig:fig2}. Panels a) and b) show the density along the $x$-axis for components $2$ and $1$, 
respectively, for different values of $N_1$ and $N_2$, keeping the total number of particles constant. It is interesting to observe 
that the size of the core of the magnetic vortex, which is of the order of the spin healing length, decreases when $N_1$ increases, 
a clear signature of the $1/\sqrt{N_1}$ dependence of $\xi_s$, as discussed for the homogeneous case. Panel c) displays the total 
density for the same values of the global polarization. For a small dip, the latter coincides with the total density profile of the 
ground state (i.e. without the vortex) of an interacting mixture due to the quasi-incompressibility of the density channel with 
respect to the spin channel. The inset in panel c) shows that as expected from the general discussion, the larger the ratio $N_2/N_1$ 
the smaller the dip in the total density. The same would occur by decreasing $\delta g$ which however will also make the vortex 
core larger and eventually comparable with the size of the trapped gas.

Finally Fig. \ref{fig:fig2}(d) shows that the magnetization change is localised within the vortex core with a maximum spin magnetization 
(at the position of the vortical axis) independent on the global polarization for a fixed total number of particles $N$. It is also 
worth to mention the fact that at distances larger than the spin healing length, but still far from the edges of the condensate, 
the magnetization is constant.

\subsection*{Stability of magnetic vortices}

\begin{figure}[t!]
\includegraphics[width=\linewidth]{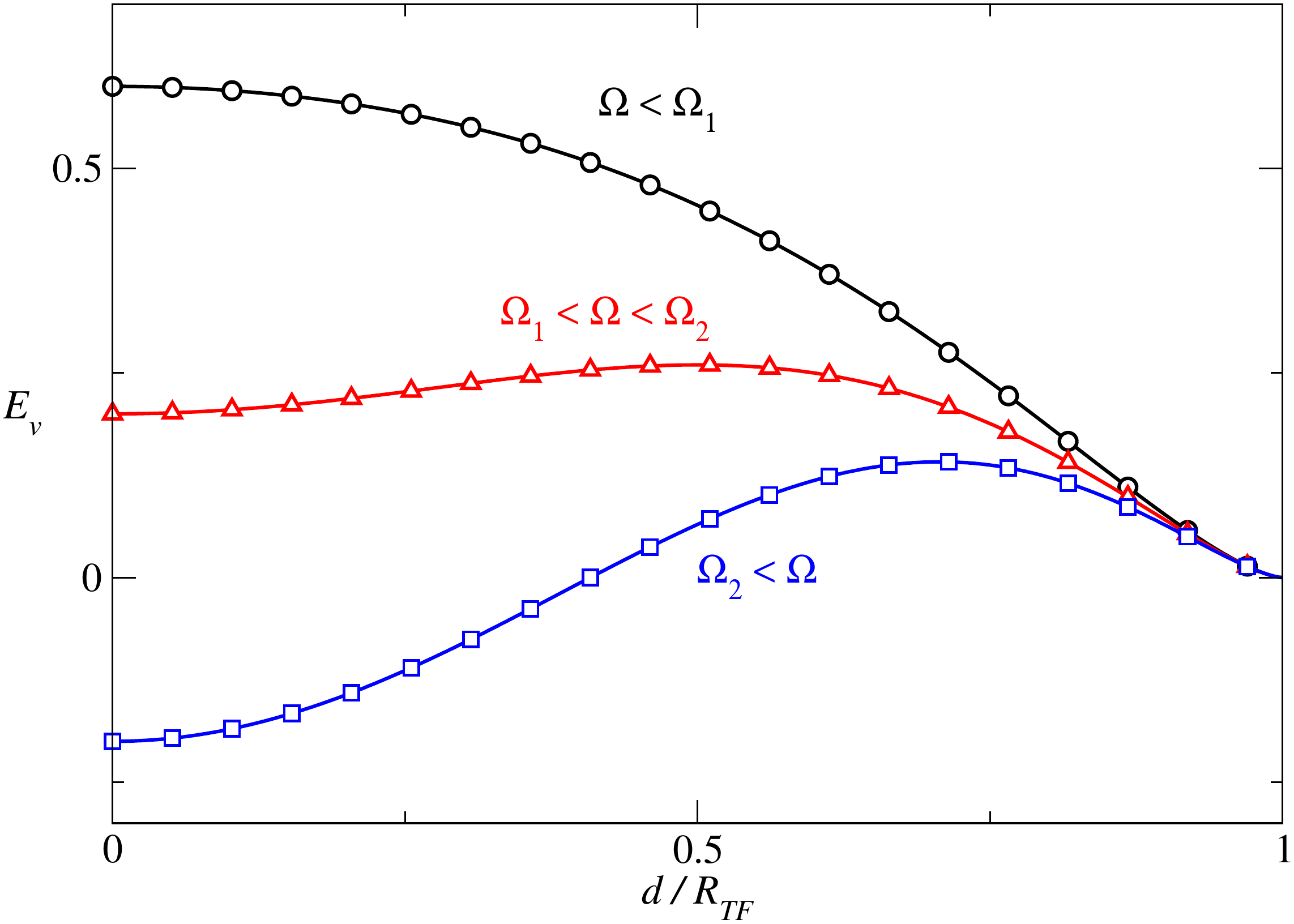}
\caption{Heuristic figure that represents the vortex energy in arbitrary units as a function of the distance $d$ (in units of the 
Thomas-Fermi radius $R_{\rm TF}$) of a vortex with respect to the center of a harmonic trap. The black line with circles represents 
the case of $\Omega<\Omega_1$ (energetic and dynamic instability), the red line with triangles the case of $\Omega_1<\Omega<\Omega_2$ 
(dynamic stability but energetic instability) and the blue line with squares the case $\Omega>\Omega_2$ (energetic and dynamic stability).}
\label{fig:fig11}
\end{figure}

In this section we will study the stability of magnetic vortices by looking at the energetic behavior of off-centered magnetic 
vortices in a frame rotating with angular velocity $\Omega$. This corresponds to adding the term $-\Omega\,\hat{L}_z$ to the hamiltonian, 
where $\hat{L}_z$ is the angular momentum operator. It is known that for single-component vortices, there exist three different 
scenarios \cite{Fetter2009,Pitaevskii2016}, shown in Fig. \ref{fig:fig11}. The first one, which appears below a certain critical 
frequency $\Omega_1$, corresponds to the case in which the vortex is nor energetically neither dynamically stable, and it will be 
pushed out of the condensate (black line with circles). The second scenario appears above this critical frequency, but below a second 
critical frequency $\Omega_2$. In this case, the vortex is dynamically stable although energetically unstable, its energy being higher 
than the value in the absence of the vortex. If the vortex is close enough to the minimum of the harmonic trap, it will remain confined 
in the center of the condensate in a metastable configuration. If, instead, it is too far from the center of the trap, it will be 
pushed away (green line with triangles). Above $\Omega_2$, the vortex is both energetically and dynamically stable, and will always 
like to stay in the center of the trap (blue line with squares). The critical angular velocity $\Omega_2$ is simply given by the 
identity $\Omega_2 = E_{MV}/L_z = E_{MV}/(N_1 \hbar)$ where we have used the value $L_z=N_1\hbar$ for the angular momentum of the 
vortex of the component $1$ located in the center of the trap. To calculate $\Omega_2$ we have considered a slightly off-centered 
magnetic vortex and identified the value of $\Omega$ at which the energy of the displaced vortex is not increasing, nor decreasing 
with respect to the value of the undisplaced vortex.

\begin{figure}[t!]
\includegraphics[width=\linewidth]{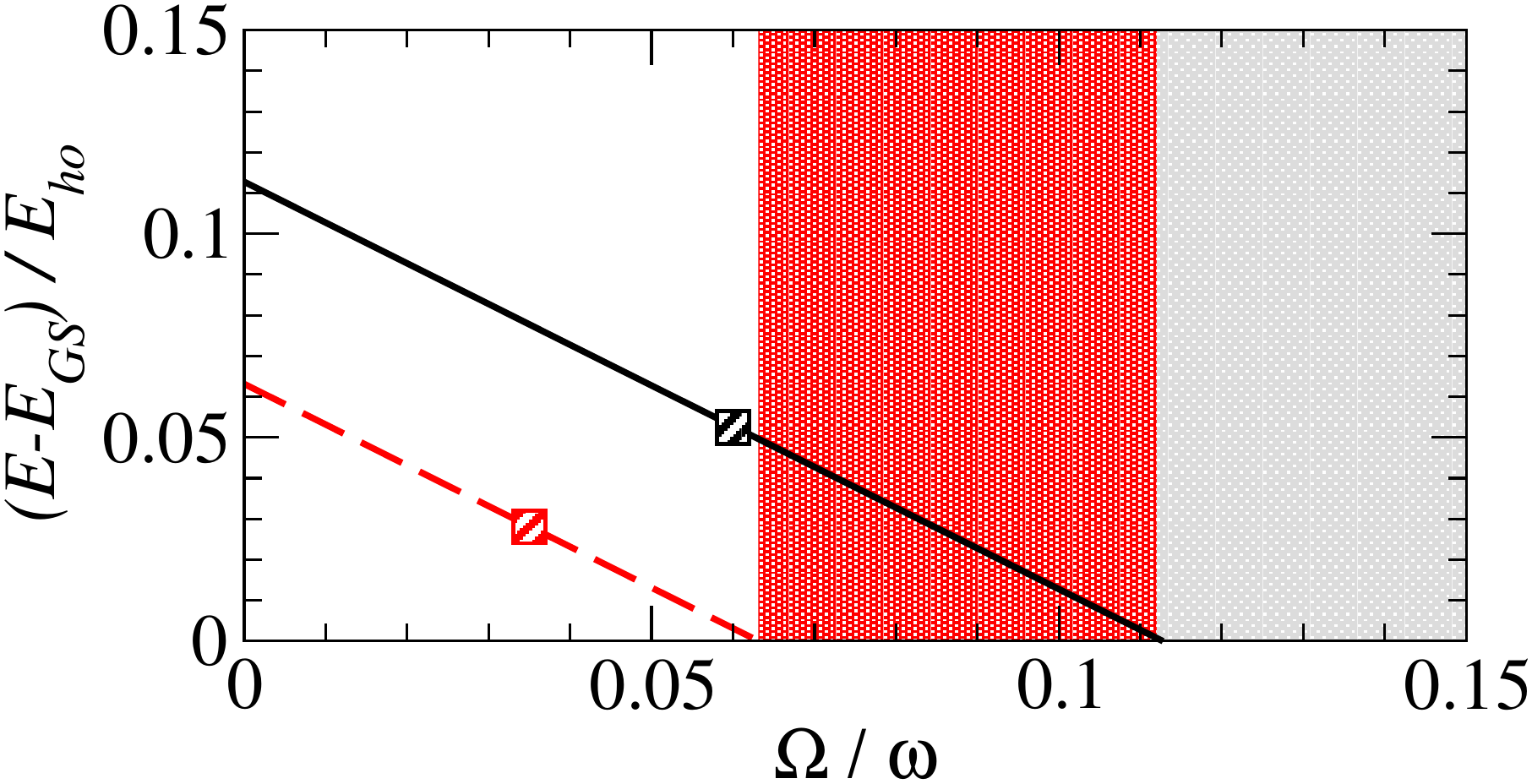}
\caption{Energy of normal vortices (solid black line) and magnetic vortices (dashed red line), with 
respect to the ground state, for different values of the rotation frequency $\Omega$. The point at 
which the two energies cross, which corresponds to the crossing with the horizontal axis, indicates 
the critical frequency at which the vortex state becomes energetically favourable with respect to the 
ground state. This ranges the region of stability of such vortices, colored in red for the region in 
which only the magnetic vortex is stable, and in grey for the region in which both normal and magnetic 
vortices are stable. There is no region in which normal vortices are stable and magnetic vortices are 
not. The squared points indicate the transition between the region of parameters in which the vortex 
is a maximum (left), and a local minimum (right), both for the normal vortex (black) and magnetic 
vortex (red). The mass and interaction parametes are those of sodium, with a harmonic frequency equal 
to $\omega=2\pi\times15.92$ Hz and $N=10^5$.}
\label{fig:fig5}
\end{figure}

Figure \ref{fig:fig5} shows our numerical results pointing out the difference between the energy of a normal (solid black line) 
and a magnetic (dashed red line) vortex and the corresponding ground state energy, as a function of the angular velocity $\Omega$. 
At a certain value of the anguar velocity $\Omega_2$, the vortex state becomes energetically favourable with respect to the 
configuration without the vortex \cite{Fetter2009} (see Ref. \cite{Aftalion2001} for the 2D case), which means that magnetic 
vortices become a global minimum in the energy landscape. The region of energetic stability is represented in the colored areas: 
the red area corresponds to values of $\Omega$ at which the magnetic vortex is  energetically stable, while the normal vortex 
is unstable. In the grey area, both vortices are energetically stable. The figure explicitly shows that magnetic vortices are 
stabilized at angular velocities smaller than in the case of normal vortices. The squares in the figure represent the critical 
value $\Omega_1$ that separates the dynamically stable and unstable regimes.

\section{From bound condensates to magnetic vortices}
\label{sect:boundto}

\begin{figure}[t!]
\includegraphics[width=\linewidth]{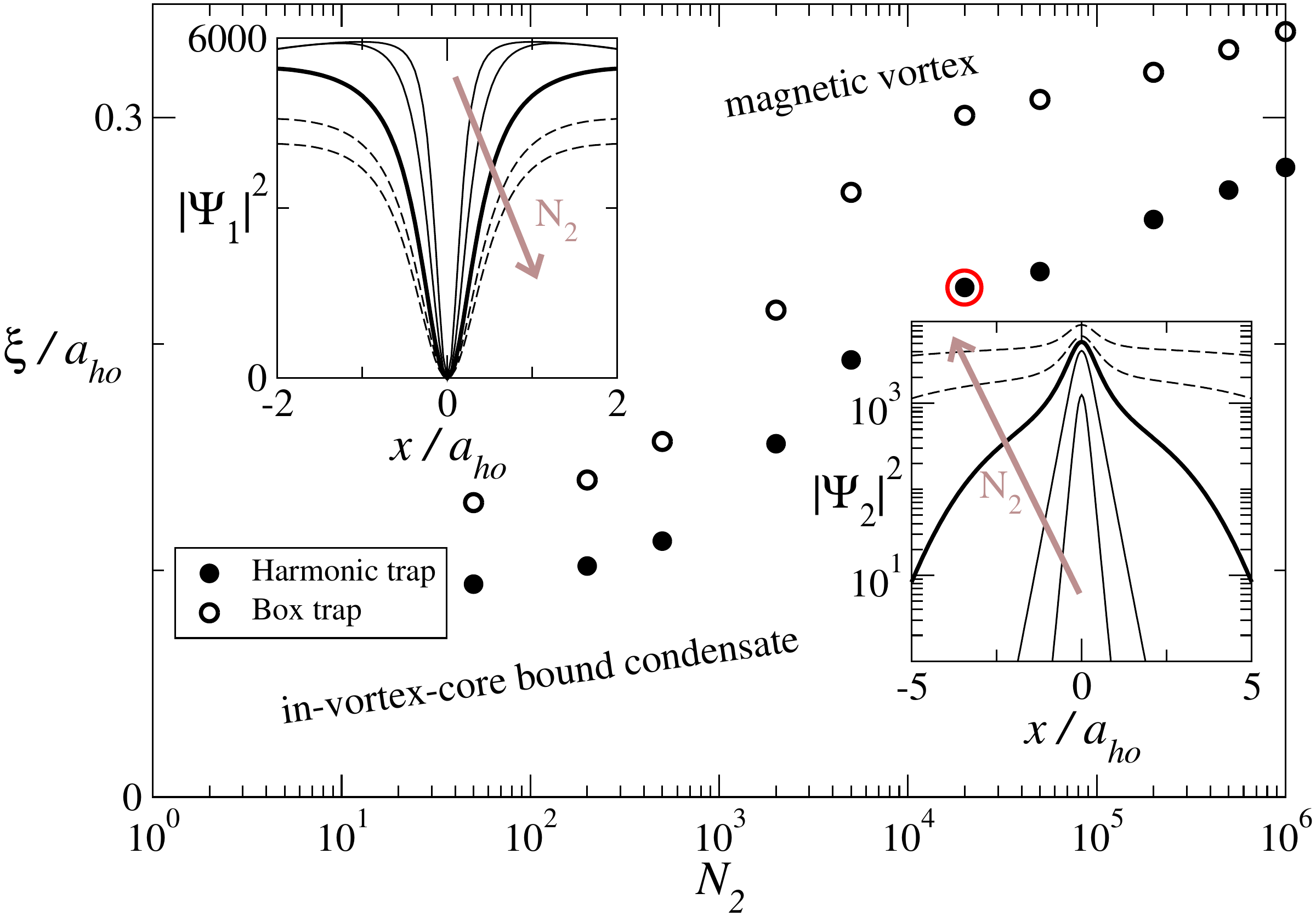}
\caption{The main figure shows the healing length of the vortex component as a function of the number 
of particles $N_2$ of the other component (black filled circles), in the case in which the system is 
harmonically trapped. The insets display the density of component $1$ (left top, in linear scale) and 
$2$ (right bottom in logarithmic scale), as a function of $x$, for different values of $N_2$, each 
curve corresponding to a different point in the main figure, in increasing order, following the brown 
arrow. The solid line corresponds to the point surrounded by a red circle. The parameters are those of 
sodium, with a trapping frequency $\omega=2\pi\times15.92$ Hz, and $N_1=10^5$. For comparison, we also 
plot in the main graph the dependence of the healing length as a function of $N_2$ for a box trap with 
radius $R=64\,\mu$m, in terms of the harmonic oscillator length (black open circles) of the first case. 
The densities are written in harmonic oscillator units.}
\label{fig:fig7}
\end{figure}

When the number of particles of the component $2$ is much smaller than the number of particles in the component $1$ hosting the 
vortex, the magnetic vortex picture fails. If we add a single particle to the component $2$, this will be in fact bound in the 
core of the vortex. However, if we add too many particles $2$, the vortex will not be able to bound all of them and component 
$2$ will diffuse outside the vortex core region eventually forming the magnetic vortex configuration. There are two main features 
that confirm this scenario. The first one is the saturation of the healing length of the vortex as $N_2\to\infty$. The saturated 
healing length is precisely the spin healing length that we found in the previous sections, which exhibits a very weak dependence 
on the density $n_{2,0}$. The second feature is the disappearence of the typical exponential tail characterizing the wave function 
of the condensate $2$ in its bound configuration.

The above effects can be clearly seen in Fig. \ref{fig:fig7}, where we report the results for the harmonic oscillator potential 
with frequency $\omega=2\pi\times15.92$ Hz, and for a box potential with radius $R=64\,\mu$m. The main figure shows how the width 
of the vortex, represented by $\xi$, increases when $N_2$ increases, to reach a saturated value with $N_2$ (notice the logarithmic 
scale in the horizontal axis), that coincides with the {\it in-medium} spin healing length $\xi_s$. The width has been calculated 
by fitting the ansatz of Eq.(\ref{eq:functionc1}) to the wave function found numerically. In the figure we have also plotted the 
density of the component $1$ (left top inset) and $2$ (right bottom inset), for different values of $N_2$. The inset on the right 
explicitly reveals the exponential decay (note the logarithmic scale in the vertical axis) of the density for small enough values 
of $N_2$, corresponding to the solid thin curves. There is actually a visible change of the decay, starting from  the solid thick 
line (corresponding  to the circled point in the main figure). For larger values of $N_2$ the clear deviations from the exponential 
decay reveal the  onset of the formation of the magnetic vortex. It is also important to observe that the ratio between the healing 
length at large and small values of $N_2$ provides a result very close to the value $\sqrt{g/2\delta g}=2.69$ predicted in Sect. 
\ref{sect:homomatter}.

A simple estimate of the value of the number of atoms $N_2$ providing the onset of the diffusive nature of the particles $2$ outside 
the vortical region and the consequent formation of the magnetic vortex, can be obtained by imposing that the average value of the 
density of the trapped condensate inside the vortex equals the density $n_{1,0}$ of the first component. Such a condition leads to 
the estimate 
$$N_2\sim N_1 \left(\frac{\xi_s}{R}\right)^D\,,$$ 
where $D$ is the dimensionality of the system and $R$ gives the size of the system (of the order of the Thomas-Fermi radius). As an 
example, the typical ratio between the spin healing length and the Thomas-Fermi radius ranges from $1/10$ to $1/50$, which yields 
a ratio between $N_2$ and $N_1$ of the order of $10^{-2}$-$10^{-4}$. These numbers are compatible with the result given in Fig. 
\ref{fig:fig7}.

\section{Magnetic dark solitons in homogeneous matter}
\label{sect:ms_homomatter}

In this Section we consider the case where the component $1$ is hosting a dark soliton. We show that assuming the incompressibility 
condition (\ref{eq:incompressibility}), it is possible to find a solution of the coupled Gross-Pitaevskii equations that exhibits 
important analogies with the case of the magnetic vortex. A peculiar feature is that in the case of solitons we are able to obtain 
systematic analytical results for all values of the polarization. The magnetic soliton for a balanced mixture was introduced in 
\cite{Qu2016}. We generalise the results for the dark soliton in the unbalanced case and show how by increasing the atom number of 
the second component one can eventually reach the solitonic solution in the incompressibility limit where the width of the soliton 
is exactly given by the renormalised healing length (\ref{ximedium}). Our result explains also the observed insensitivity with respect 
to the polarisation in the emergence of magnetic-like solitons, as reported in the recent experiment carried out in Ref. \cite{Danaila2016}.

The magnetic dark soliton is obtained by considering the one dimensional version of Eq. (\ref{GPEnergy-const}). Let us 
consider a soliton at rest along the $z$ direction with the soliton plane at $z=0$, and $\phi_i=0$. We can use the Ansatz 
$\sqrt{n_{1,0}}f_1(z)=\sqrt{n}\cos{(\theta(z)/2)}$ in Eq.(\ref{GPEnergy-const}) to obtain the expression:
\begin{align}
\label{eq:eninc}
\frac{\varepsilon_{MS}(\theta)}{\delta g n^2}=\frac{1-p}{4}(\partial_\eta \theta)^2+\cos^2(\theta/2)(\cos^2(\theta/2)-1+p)\,
\end{align}
for the energy density, where we have defined the polarisation $p=(n_{2,0}-n_{1,0})/(n_{2,0}+n_{1,0})$. By minimizing the energy with 
respect to the function $\theta(z)$, one finds the following differential equation:
\begin{align}
\label{difftheta}
\partial_\eta^2 \theta+\sin(\theta)\frac{\cos(\theta)+p}{1-p}=0\,,
\end{align}
which admits the ground state uniform solution (absence of the soliton), setting $\theta(z) =0$. Equation (\ref{difftheta}) also admits 
a non trivial solitonic solution yielding the result 
\begin{align}
n_1(z)=n_{1,0}\frac{\cosh(\sqrt{1+p}\, z/\xi_s)-1}{\cosh(\sqrt{1+p}\, z/\xi_s)+p}
\end{align}
for the density of the component $1$ hosting the soliton, for any value of the polarization $p$. In the case $n_{1,0}=n_{2,0}=n/2$, i.e., 
$p=0$, the solution reduces to the magnetic dark soliton solution of Ref. \cite{Qu2016}, while for $p\rightarrow 1$ gives the Tsuzuki 
solution \cite{Tsuzuki1971}, with the rescaled value $\xi_s$ for the spin healing length, accounting for the interaction with the 
component $2$.

In conclusion, as expected, both the vortex and the soliton exhibit a similar behaviour as a function of the polarisation. In the 
$p\rightarrow 1$ case the topological objects are modified by the medium via a simple renormalisation of the healing length. In the 
opposite limit $p\rightarrow -1$ the incompressibility condition cannot be satisfied and, similarly to the case of the vortex discussed 
in the previous sections, the limiting case corresponds to  an impurity trapped by the core of the solitonic component $1$. In the 
case of the soliton such a limit can be solved exactly, since it is equivalent to a particle bound in a P\"oschl-Teller potential, 
a problem addressed in Refs. \cite{Antezza2007,Charalampidis2015}. In our case, due to the condition $0<g_{12}<g$, we find that the 
potential admits only a single bound state. The crossover between localised atoms in the soliton core and the magnetic soliton occurs 
in the same fashion as already described for the vortex.

\section{Vortices with asymmetric coupling constants}
\label{sect:bouyancy}

\begin{figure}[]
\includegraphics[width=0.8\linewidth]{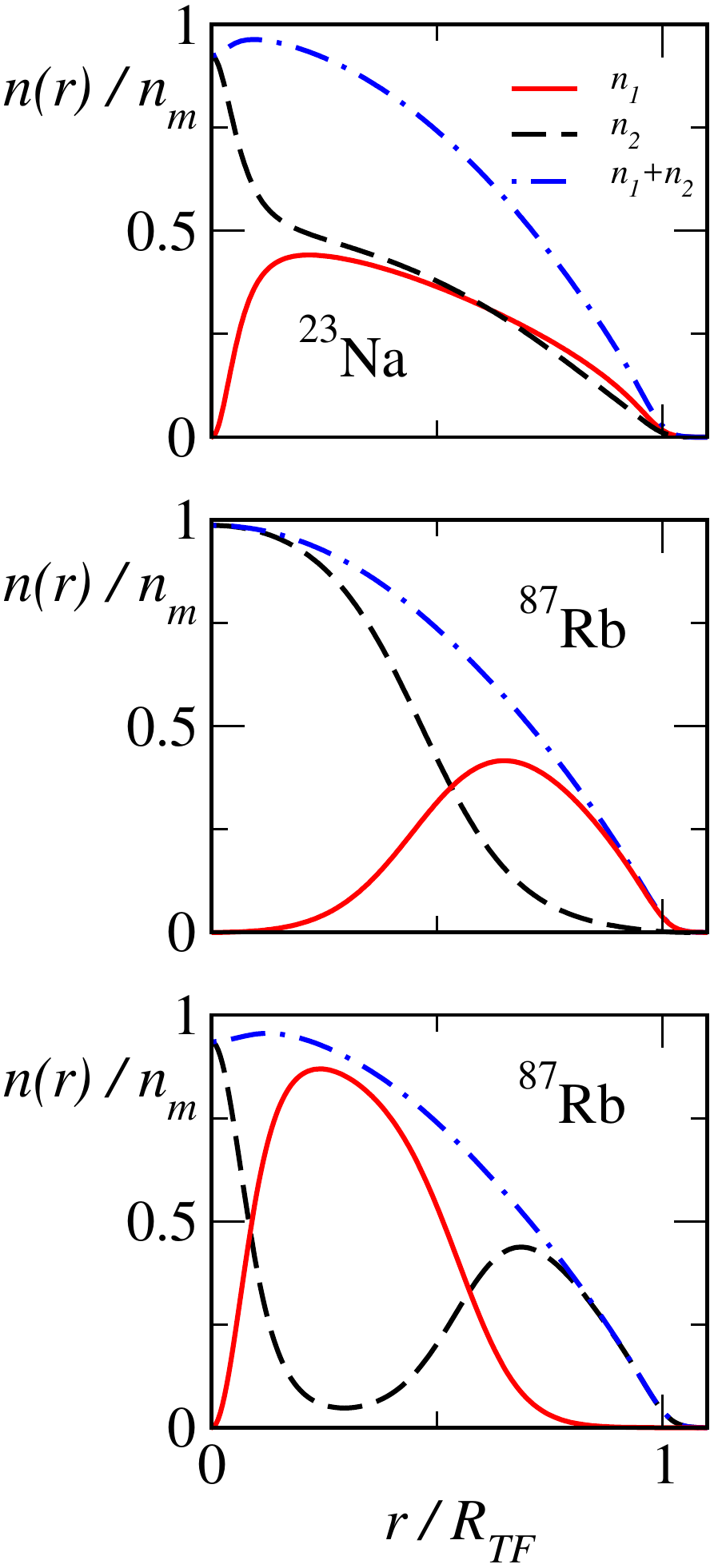}
\caption{The solid red (dashed black) line represents, density along the radial axis of component $1$ ($2$). 
The dot-dashed blue line shows the total density of the magnetic vortex configuration along the radial axis. 
In the top panel: \mbox{$a_{11}=54.54\,a_B$}, \mbox{$a_{22}=54.54\, a_B$}, and \mbox{$a_{12}=50.78\,a_B$}. 
In the middle panel: \mbox{$a_{11}=100.44\,a_B$}, \mbox{$a_{22}=95.47\, a_B$}, and \mbox{$a_{12}=98.09\,a_B$}. 
In the bottom panel: \mbox{$a_{11}=95.47\,a_B$}, \mbox{$a_{22}=100.44\, a_B$}, and \mbox{$a_{12}=98.09\,a_B$}. 
The rest of the parameters are: $N=10^5$ (total number of particles), $\omega=2\pi\times15.92$ Hz and the mass 
is that of $^{23}$Na or $^{87}$Rb, depending on the case. The densities are written in harmonic oscillator units.}
\label{fig:fig10}
\end{figure}

The magnetic vortices discussed in the main text are exhibited by miscible mixtures satisfying the condition 
$g_{11}=g_{22} > g_{12}$. The hyperfine states $|F=1,m_{\rm F}=\pm1\rangle$ of sodium satisfy this condition 
and are consequently well suited to expore experimentally the main features discussed in our paper. When the 
condition $g_{11}=g_{22} $ is not satisfied, as happens for example in the case of the two hyperfine states 
$\vert F$=$1, m_F$=$-1\rangle$ and $\vert F$=$2, m_F$=$1\rangle$  employed in \cite{Matthews1999} to generate 
vortical configurations, the resulting scenario changes in a deep way even if the miscibility condition 
$g_{12}< \sqrt{ g_{11}g_{22}}$ is satisfied, because of the occurrence of buoyancy in the presence of an 
external harmonic trap which causes phase separation between the two atomic species. This effect was actually 
observed  in Ref.  \cite{Matthews1999}, and soon after theoretically explained  by Refs. \cite{Jezek2001,Chui2001,PerezGarcia2000,Jezek2005}.

In the top and middle panels of Fig. \ref{fig:fig10} we compare the density profiles calculated in the case of 
$^{23}$Na, and already discussed in the main text (top panel), with the density profiles calculated in $^{87}$Rb, 
calculated  by imposing a vortex in the component $1$, here identified with the state $\vert F$=$1, m_F$=$-1\rangle$, 
the component $2$ corresponding to the state $\vert F$=$2, m_F$=$1\rangle$ (middle panel). The two states of 
$^{87}$Rb have asymmetric coupling constants given by \mbox{$a_{11}=100.44\,a_B$}, \mbox{$a_{22}=95.47\, a_B$} 
and \mbox{$a_{12}=98.09\,a_B$} \cite{Harber2002,Egorov2013}. They satisfy the miscibility condition, but yield 
buoyancy. Actually, even the tiny difference between $g_{11}$ and $g_{22}$ is responsible for phase separation 
in the presence of harmonic trapping, as clearly shown by the figure. In particular, while the total density 
looks similar to the case of $^{23}$Na and is scarcely affected by the presence of the quantum defect, the spin 
density  $n_1-n_2$ exhibits a very different behavior, the component $2$ providing a core pushing the rotating 
component $1$ towards the peripherical region. In this case the width of the vortex is not given by the healing 
length, but is rather fixed by the Thomas-Fermi radius of the component $2$.

It is worth mentioning that the above scenario occurs because we put the vortex in the component with the larger 
intraspecies coupling: both the bouyancy effect and the vortex state favour the second non-rotating component to 
be in the center of the trap. One could indeed also consider to put the vortex state in the component with the 
smaller intraspecies interaction. In this case the competition between the bouyancy effect and the presence of 
the vortex core leads to the alternating density configuration reported in Fig. \ref{fig:fig10} (bottom panel).

\section{Conclusions}
\label{sect:conclusions}

In this work we have studied a mixture of two Bose-Einstein condensates in which one of them holds a topological 
defect. In particular, we have focused on incompressible configurations where the total density of the system is 
not affected by the presence of the quantum defect, which then exhibits a typical magnetic nature, characterized 
by a pronounced  local magnetization. By assuming incompressibility, we have derived an exact equation chracterized 
by a length scale that can be identified with a {\it in-medium} spin healing length, fixed by the difference between 
the intraspecies and interspecies coupling constants. We have applied this equation to both magnetic vortices and 
static magnetic solitons. For the case of magnetic vortices, we have been able to obtain a numerical solution in 
the homogeneous case as well as in the trapped case, and we have found that magnetic vortices are energetically 
more stable than normal vortices.

We have also seen that when the number of particles in the component without the topological defect becomes small, 
at a certain point the incompressibility condition becomes a very bad approximation, and magnetic vortices can not 
be obtained. In the limit of few particles in component $2$ one finds bound states localized in the core of the 
topological defect. In the case of the vortex, we have explicitly analyzed the transition from few to many particles 
in component $2$, and we have pointed out that the transition between bound configurations inside the vortex core 
and diffused configurations outside the vortex core is fixed by the point where the interaction energy exceeds the 
chemical potential of the vortical component \footnote{Although the nature is completely different, it is worth 
mentioning a recent work that has predicted a transition in vortex-core structures in Bose-Fermi superfluids 
\cite{Pan2017}.}.

When the two intraspecies interactions are not equal the system exhibits buoyancy in the presence of harmonic 
trapping and we have explicitly investigated also this case that was relevant, for example, in the paper of Ref. 
\cite{Matthews1999} devoted to the first experimental realization of quantized vortices.

We have also addressed the problem of the magnetic soliton, which is exactly solvable for all values of the polarization. 
We have seen that in the balanced case, the results of Ref. \cite{Qu2016} are recovered while, in the case where the 
component cotaining the vortex is a minority, one recovers the Tsuzuki solution \cite{Tsuzuki1971} with a renormalized 
spin healing length. 

The experimental realisation of magnetic vortices discussed in our work could be done using the same procedure as 
in Ref. \cite{Matthews1999}. This technique is based on the microwave transfer carrying angular momentum between 
two hyperfine levels, and should be applied to the states $|F=1,m_{\rm F}=\pm1\rangle$ of sodium atoms.

\section*{Acknowledgments}

This project has received funding from the European Union's Horizon 2020 research and innovation programme under grant 
agreement No. 641122 "QUIC". We would also like to thank enlightening discussions with A. Fetter, and the members of 
the BEC Center. We are also indebted to M. Pi, for his ideas to numerically solve the problem in a homogeneous system.

\bibliography{bibtex.bib}

\end{document}